# Visibility of dichalcogenide nanolayers

M.M. Benameur[1], B. Radisavljevic[1], S. Sahoo[1], H. Berger[2], A. Kis[1]

[1]*Electrical Engineering Institute, School of Engineering, EPFL, 1015 Lausanne, Switzerland*

[2]*Institute of Condensed Matter Physics, School of Basic Science, EPFL, 1015 Lausanne, Switzerland*

**Dichalcogenides with the common formula $MX_2$ are layered materials with electrical properties that range from semiconducting to superconducting. Here, we describe optimal imaging conditions for optical detection of ultrathin, two-dimensional dichalcogenide nanocrystals containing single, double and triple layers of $MoS_2$, $WSe_2$ and $NbSe_2$. A simple optical model is used to calculate the contrast for nanolayers deposited on wafers with varying thickness of $SiO_2$. The model is extended for imaging using the green channel of a video camera. Using AFM and optical imaging we confirm that single layers of $MoS_2$, $WSe_2$ and $NbSe_2$ can be detected on 90nm and 270 nm $SiO_2$ using optical means. By measuring contrast under broad-band green illumination we are also able to distinguish between nanostructures containing single, mono and triple layers of $MoS_2$, $WSe_2$ and $NbSe_2$.**

The family of transition metal dichalcogenides with the common formula $MX_2$ where M stands for transition metals (M=Mo, W, Nb, Ta, Ti) and X for chalcogens (Se, S or Te) displays a rich variety of physical properties. Depending on the metal and the chalcogen involved, their electrical properties span the range from semiconducting to superconducting. Bulk dichalcogenide crystals are composed of vertically stacked layers bound together by weak van der Waals interaction. Just as in the case of graphene[1], single dichalcogenide layers can be extracted from bulk crystals[2-3] and deposited on substrates for further studies. Single $MX_2$ layers present a wide range of systems for studying mesoscopic transport in 2D and could find practical applications complementary to those of graphene. Bulk $WSe_2$ has for example been used in past for fabrication of photovoltaic cells[4], whereas $MoS_2$ nanotubes[5] and nanowires[6] show confinement effects in their electronic and optical properties. Semiconducting dichalcogenides could also be interesting for fabrication of nanoscale field effect transistors[7-8] while superconducting $NbSe_2$ could be a model for studying superconductivity in low-dimensional systems at mesoscopic scales.[9-10]

Locating and identifying single nanolayers of materials such as graphite[1] or semiconducting transition metal dichalcogenides[3] such as $MoS_2$ or $WSe_2$ is the first, enabling step in the study and practical applications of these materials. Atomic force microscopy (AFM) can be used to accurately determine both the vertical and lateral dimensions of nanolayers deposited on insulating substrates such as $SiO_2$. AFM imaging is however time-consuming and the relatively slow throughput of the technique is a serious drawback. Scanning electron microscopy (SEM) or transmission electron microscopy (TEM) could also be used here, but contamination due to electron beam-induced deposition or knock-on damage in TEM could be a serious problem here.

Optical imaging offers the possibility of simple, rapid and non-destructive characterization of large-area samples. In the case of graphene deposited on $SiO_2$, it has been found that even the presence of a single layer can produce a detectable contrast with respect to the interference color of the underlying oxide layer.[11-12] The simplicity and accessibility of this detection scheme was one of the most important factors that allowed the rapid spread in graphene-related research.

It is however not clear what would be the optimal conditions for optical detection of dichalcodenide nanolayers. It could even be possible that such nanolayers deposited on 300nm $SiO_2$, commonly used for graphene-related studies, might be invisible because of a particularly unfortunate set of interference conditions.

We have therefore decided to calculate the contrast for several different types of nanolayers deposited on $SiO_2$ in order to determine the optimal imaging conditions for their optical detection. In this work we focus on three representative dichalcogenide materials that might be most interesting for future studies: semiconducting $MoS_2$ and $WSe_2$ that could be useful for fabrication of nanoscale field effect transistors[7-8] and superconducting $NbSe_2$ which could be a new model for studying superconductivity in low-dimensional systems.[9-10]

In analogy with graphene[11-14], the contrast between dichalcogenide nanolayers such as the one depicted on Figure 1., and the underlying $SiO_2$ substrate is due to phase shift of the interference color and material opacity. In order to calculate this contrast, we consider the stacking of two thin films (2D dichalcogenide material and $SiO_2$) on top of a third semi-infinite film (degenerately doped n-type Si), as depicted on Figure 1. The 2D nanolayer is modeled as a thin homogeneous film of thickness $d_1$ with complex refractive index $n_1$ where $Re(n_1)$ is the optical refractive index and $-Im(n_1)$ is the absorption coefficient. Previously published values for the refractive indices and absorption coefficients of bulk $MoS_2$, $WSe_2$ and $NbSe_2$ are available in the literature[15-17]. The $SiO_2$ layer of thickness $d_2$ is optically characterized by a wavelength dependent refractive index $n_2(\lambda)$ with only a real part[18], ranging from 1.47 at 400nm to 1.455 at 700nm. As the thickness of the degenerately doped Si layer (525µm) is several orders of magnitudes larger than the corresponding skin depth, it can be considered as a semi-infinite film. For normal light incidence, the intensity of reflected light from a stacking of two thin films on top of a semi-infinite layer is given by[11, 19]

$$R(n_1) = \left| \frac{r_1 e^{i(\phi_1+\phi_2)} + r_2 e^{-i(\phi_1-\phi_2)} + r_3 e^{-i(\phi_1+\phi_2)} + r_1 r_2 r_3 e^{i(\phi_1-\phi_2)}}{e^{i(\phi_1+\phi_2)} + r_1 r_2 e^{-i(\phi_1-\phi_2)} + r_1 r_3 e^{-i(\phi_1+\phi_2)} + r_2 r_3 e^{i(\phi_1-\phi_2)}} \right|^2 \quad (1)$$

where

$$r_1 = \frac{n_0 - n_1}{n_0 + n_1}, \quad r_2 = \frac{n_1 - n_2}{n_1 + n_2}, \quad r_3 = \frac{n_2 - n_3}{n_2 + n_3} \quad (2)$$

are the relative indices of refraction and $\phi_i = \frac{2\pi d_i n_i}{\lambda}$ are the phase shifts induced by changes in the optical path.

On the other hand, the reflected light intensity in the absence of a nanolayer can be found by substituting $n_1=1$:

$$R(n_1 = 1) = \left| \frac{r_2' e^{i(\phi_2)} + r_3 e^{-i(\phi_2)}}{e^{i(\phi_2)} + r_2' r_3 e^{-i(\phi_2)}} \right|^2 \quad (3)$$

where $r_2' = \frac{n_0 - n_2}{n_0 + n_2}$ is the relative index of refraction at the interface between air and the dielectric thin film.

The contrast is defined as the relative intensity of reflected light in presence and absence of the 2D dichalcogenide material and can be written as:

$$Contrast = \frac{R(n_1 = 1) - R(n_1)}{R(n_1 = 1)} \tag{4}$$

In order to determine optimal conditions for the optical detection of nanolayers we plot the calculated contrast as a function of incident light wavelength and $SiO_2$ thickness on Figure 2. For all three materials and $SiO_2$ thickness lower than 300nm, the contrast for visible light wavelengths exhibits two characteristic bands with high, positive contrast. They roughly correspond to $SiO_2$ thickness in the 50-100 nm and 200-300 nm range, implying that dichalcogenide nanolayers should in principle be visible on substrates with such oxide thicknesses for at least some spectral ranges of the visible light. In the 100-150nm $SiO_2$ thickness range, we expect to see weaker, negative contrast for red light illumination.

In the next step, we generalize the model for broadband illumination, making it possible to model contrast values observed with standard color cameras, avoiding the need for additional color filters. We can compute the effective contrast by calculating the average contrast weighed by the camera response function $S(\lambda)$ for a given channel (red, green or blue). The response function is available in technical specifications for a given camera and is primarily determined by the Bayer filter in front of the camera's CCD, implying that our findings are relevant to color cameras from other manufacturers. We limit ourselves to the green channel only (495nm-530nm) as the typical Bayer filter used in color cameras contains 50% green and only 25% of red and blue elements each.

The contrast in the green channel is then given by:

$$Contrast_{green}(d_{substrate}) = \frac{\int_{\lambda=495nm}^{\lambda=530nm} S(\lambda) Contrast(\lambda, d_{substrate}) d\lambda}{\int_{\lambda=495nm}^{\lambda=530nm} S(\lambda) d\lambda} \quad (5)$$

Calculated values are reported on Figure 3. For all the three materials that we studied, we find two characteristic peaks in the 0-300nm region due to constructive interference. They are located at 78nm and 272nm in the case of $MoS_2$, at 80nm and 274nm for $WSe_2$ and 84nm and 274nm for $NbSe_2$. For substrate thicknesses ranging from 100nm to 200nm the model shows minimum contrast values indicating that under these conditions monolayers would appear to be very faint or even invisible.

Based on these calculations, we predict that substrates with $SiO_2$ thicknesses of 90nm and 270nm should result in sufficient contrast for optical detection of dichalcogenide nanolayers. We note that these values are close to optimal conditions for imaging graphene (90nm and 280nm).[11]

We proceed by depositing individual dichalcogenide nanolayers on substrates with 90, 250 and 270nm $SiO_2$ thickness using the mechanical exfoliation technique commonly used for graphene deposition.[1] Briefly, we attach a piece of scotch tape to the surface of a bulk crystal. The tape is peeled off together with microscopic fragments of the desired material. It is then rubbed across a $SiO_2$ surface, resulting in mechanical exfoliation of nanolayers that are readily identified in the debris using an optical microscope. In this study we used naturally occurring $MoS_2$ (SPI supplies) as well as high-quality $WSe_2$ and $NbSe_2$ crystals grown in-house using the vapor transport method.

After mechanical exfoliation, we image the surface of the sample using an optical microscope (Olympus BX51M) equipped with a color camera (AVT Pike F-505C). After having located the nanolayers with lowest contrast values using the optical microscope, we image the sample using an atomic force microscope (Asylum Research Cypher) in order to measure nanolayer height using AC-mode imaging. Representative optical and AFM images are shown on Figure 4. Based on AFM imaging, we measure the following thicknesses: 6.75 Å for $MoS_2$, 6.7 Å for $WSe_2$ and 6.86 Å for $NbSe_2$. These values correspond well to interlayer separation in dichalcogenide crystals, proving that we have managed to exfoliate single layers. Corresponding profiles of optical contrast reported on Figure 4. show contrast values for single layer of $MoS_2$, $WSe_2$ and $NbSe_2$ in the 25-30% range in the green channel. In the case of $WSe_2$ and $NbSe_2$, our values show excellent agreement with calculations shown on Figure 3. In the case of $MoS_2$, the measured contrast value at 270nm shows a significant discrepancy with respect to the model. Hoping to improve the accuracy of our model we attempted to refine it by considering a very thin layer of water adsorbed between the nanolayer and the substrate. This assumption however did not lead to more accurate results. In fact, as water has a very small extinction coefficient, no additional absorption of light takes place and the addition of a water layer only adds a phase factor proportional to its thickness. The observed discrepancy between calculated and observed values of contrast might be due to a variation of optical properties of $MoS_2$ with layer number, warranting further studies of optical properties of ultrathin $MoS_2$. We have also used the AFM ascertain the thicknesses of "darker" flakes presumably containing multiple layers. We find that the observed contrast increases with the number of layers, as shown on Figure 5. The difference in contrast between double and triple layer structures is sufficient to distinguish between them using optical imaging only.

To summarize, we have calculated the expected contrast between thin layers of $MoS_2$, $WSe_2$ and $NbSe_2$ dichalcogenide crystals and the underlying $SiO_2$ substrate. Contrast in the band corresponding to green light (495-530nm) is maximized for all three materials using 90nm and 270nm oxide layer thicknesses. Mechanical exfoliation followed by optical and AFM imaging has confirmed that single and multilayer dichalcogenide nanostructures can be visualized on substrates with proposed oxide thicknesses with easy differentiation between structures containing single, double and triple layers. Optical imaging can therefore be used as a rapid, non-invasive and low cost method for the detection of dichalcogenide nanolayers, paving the way for further studies of these nanomaterials.

## ACKNOWLEDGEMENTS

Substrate preparation was carried out in EPFL Center for Micro/Nanotechnology (CMI). This work was financially supported by ERC grant no. 240076 and Swiss SNF grant no. 200021_122044. Single crystals of $WSe_2$ and $NbSe_2$ were grown at the EPFL Crystal growth facility, financially supported by NCCR MANEP.

# FIGURES

Table of contents figure

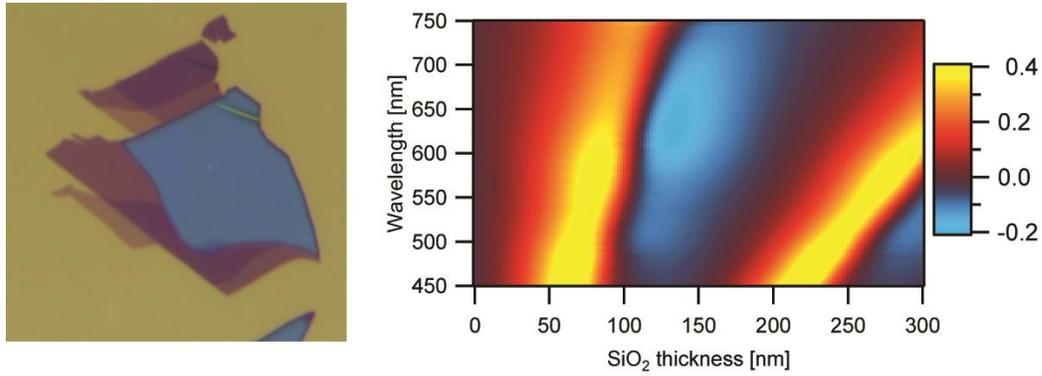

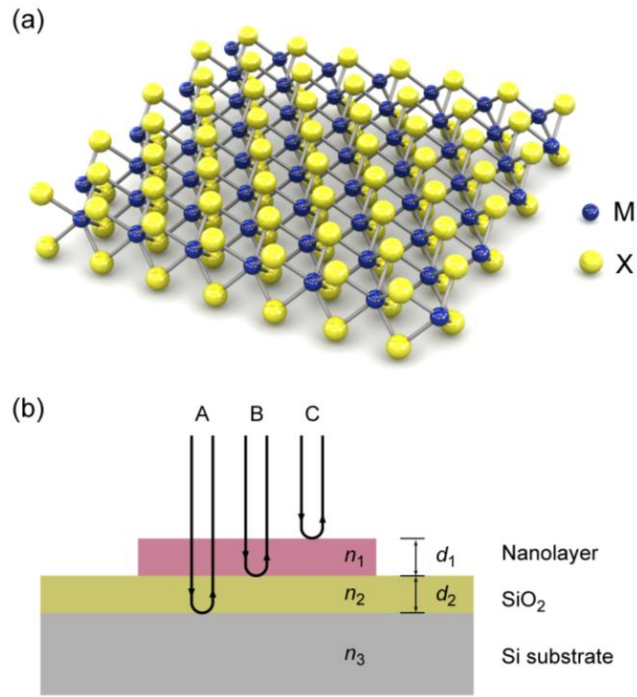

**Figure 1.** (a) Three-dimensional representation of a dichalcogenide monolayer with a generic formula $MX_2$ (b) Schematic depiction of optical reflection and transmission for nanolayer with thickness $d_1$ and complex index of refraction $n_1$ deposited on a $SiO_2$ layer characterized by thickness $d_2$ and index of refraction $n_1$ that is grown on top of a degenerately doped Si substrate. Nanolayes deposited on $SiO_2$ are visible due to interference between light rays A, B and C reflected at various interfaces in the stack.

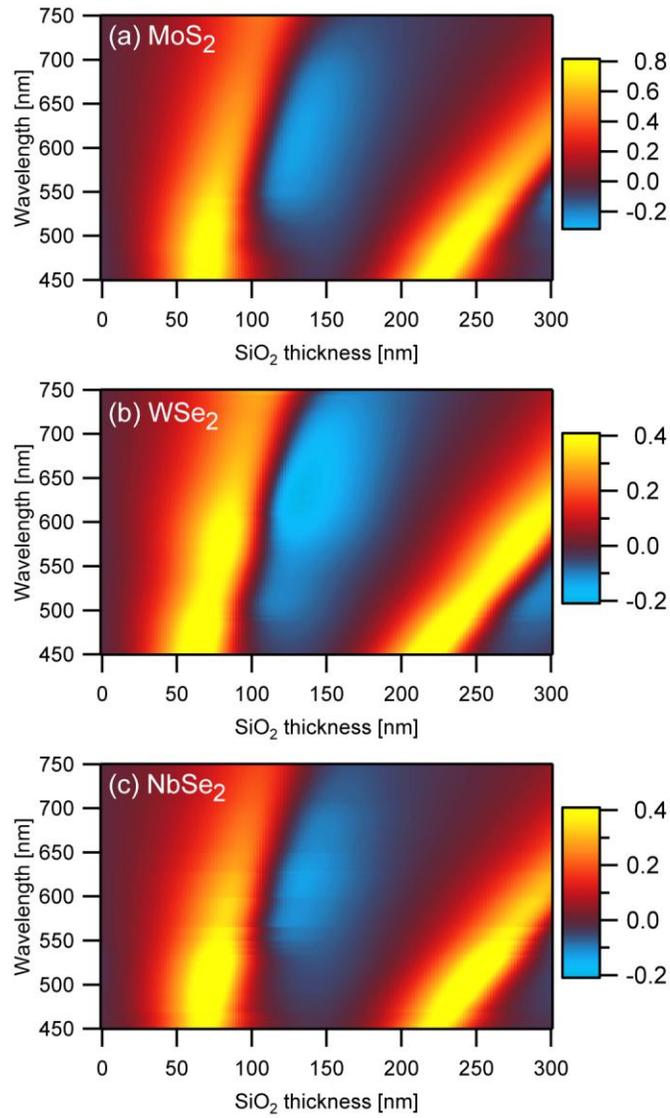

**Figure 2.** Color plot of calculated contrast as a function of incident light wavelength and $SiO_2$ layer thickness for (a) $MoS_2$ (b) $WSe_2$ and (c) $NbSe_2$. Dichalcogenide nanolayes are expected to be visible on substrates with oxide thickness in the 50-100 nm and 200-300 nm range. In the 100-150nm $SiO_2$ thickness range, we expect to see weaker, negative contrast for red light illumination.

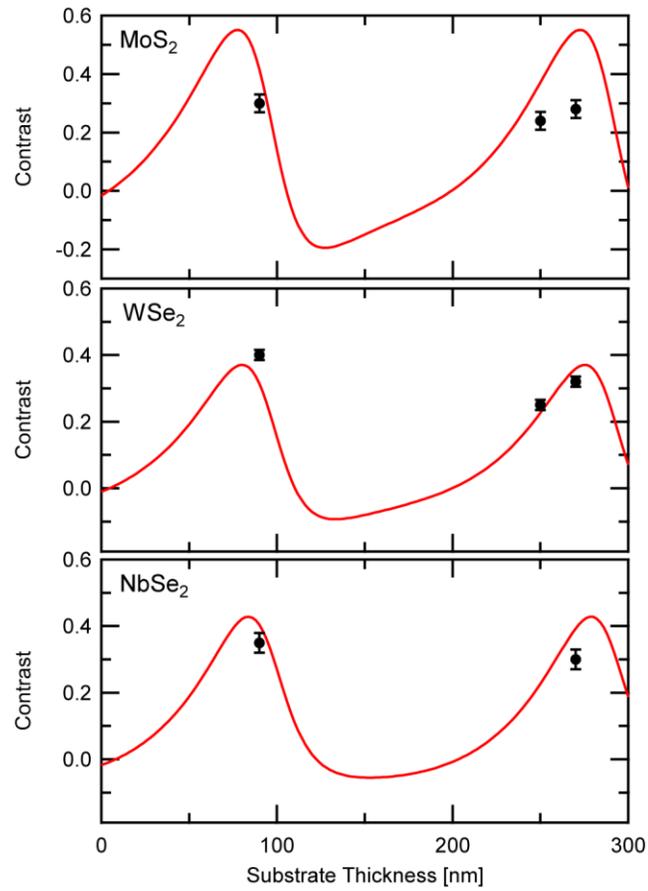

**Figure 3.** Calculated contrast values for $MoS_2$, $WSe_2$ and $NbSe_2$ deposited on $SiO_2$ substrates with varying thickness. Curves represent contrast for broadband illumination and detection using the green channel (495-530nm) of a color camera. Black dots are experimental data points.

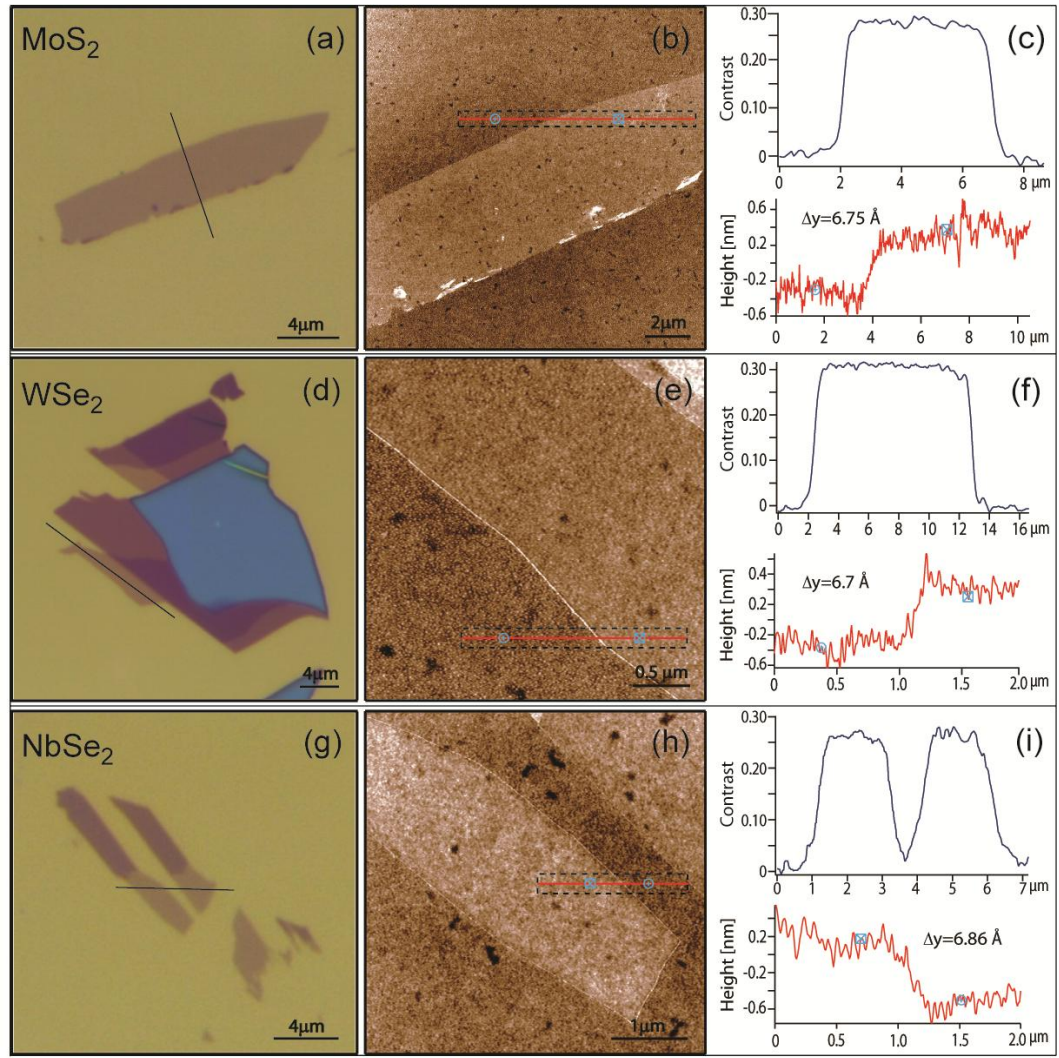

**Figure 4.** Optical and AC-mode AFM images of dichalcogenide nanolayers deposited on 270nm $SiO_2$ with corresponding contrast and height profiles of monolayers: (a - c) for $MoS_2$, (d - f) for $WSe_2$ and (g - i) for $NbSe_2$. Contrast and height profiles of monolayer flakes are taken across the black lines drawn on optical images, and red lines on AFM images. Measured thicknesses correspond well with interlayer distances in dichalcogenide crystals. Observed optical contrast is in the 25-30% range for all three materials and is slightly lower that the values predicted in the model.

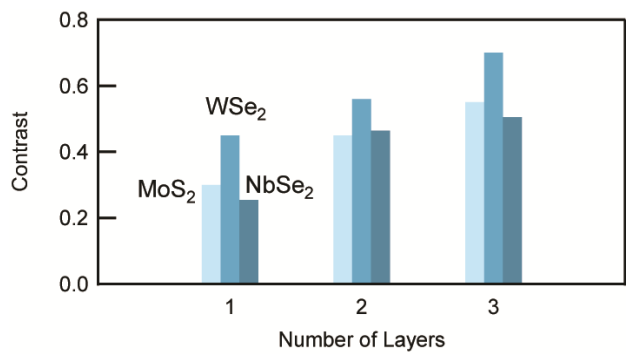

**Figure 5**. Contrast measured for $MoS_2$, $WSe_2$ and $NbSe_2$ flakes deposited on 270nm $SiO_2$ and containing different number of layers identified using AFM. For all three materials, the contrast increases with increasing layer number, indicating that optical imaging can be used to distinguish flakes with differing numbers of layers.